\documentclass[traditabstract]{aa}
\usepackage{txfonts,natbib,graphicx}
\bibpunct{(}{)}{;}{a}{}{,}

\def\ltsima{$\; \buildrel < \over \sim \;$}
\def\simlt{\lower.5ex\hbox{\ltsima}}
\def\gtsima{$\; \buildrel > \over \sim \;$}
\def\simgt{\lower.5ex\hbox{\gtsima}}
\def\fullsrc{IGR J17480-2446}
\def\src{J17480}

\def\rxte{\it RXTE}
\def\fullrxte{\it Rossi X-ray Timing Explorer}

\begin{document}

\title{The spin and orbit of the newly discovered pulsar {\fullsrc}}

\author{A.~Papitto\inst{\ref{inst1}} \and A.~D'A\`i\inst{\ref{inst2}} \and S.~Motta\inst{\ref{inst3}} \and A.~Riggio\inst{\ref{inst1}}$^{,}$\inst{\ref{inst4}} \and L.~Burderi\inst{\ref{inst1}} \and T.~Di Salvo\inst{\ref{inst2}} \and T.~Belloni\inst{\ref{inst3}} \and R.~Iaria\inst{\ref{inst2}}}

\institute{Dipartimento di Fisica, Universit\'a degli Studi di Cagliari, SP Monserrato-Sestu, KM 0.7, 09042 Monserrato, Italy\label{inst1}
\and
Dipartimento di Scienze Fisiche ed Astronomiche, Universit\'a di Palermo,
 via Archirafi 36, 90123 Palermo, Italy\label{inst2}
\and
INAF - Osservatorio Astronomico di Brera, via E. Bianchi 46, 23807 Merate, Italy
\label{inst3}
\and
INAF - Osservatorio Astronomico di Cagliari, Poggio dei Pini, Strada 54, 09012 Capoterra (CA), Italy\label{inst4}
}

\abstract{ 

We present an analysis of the spin and orbital properties of the newly
discovered accreting pulsar {\fullsrc}, located in the globular
cluster Terzan 5.  Considering the pulses detected by the {\it Rossi
  X-ray Timing Explorer} at a period of 90.539645(2) ms, we derive a
solution for the 21.27454(8) hr binary system. The binary mass
function is estimated to be 0.021275(5) $M_{\odot}$, indicating a
companion star with a mass larger than 0.4 M$_{\odot}$. The X-ray
pulsar spins up while accreting at a rate of between 1.2 and
1.7$\times10^{-12}$ Hz s$^{-1}$, in agreement with the accretion of
disc matter angular momentum given the observed luminosity.

We also report the detection of pulsations at the spin period of the
source during a {\it Swift} observation performed $\sim 2$ d before
the beginning of the {\rxte} coverage. Assuming that the inner disc
radius lies in between the neutron star radius and the corotation
radius while the source shows pulsations, we estimate the magnetic
field of the neutron star to be within $\sim2\times10^8$ G and
$\sim2.4\times10^{10}$ G. From this estimate, the value of the spin
period and of the observed spin-up rate, we associate this source with
the still poorly sampled population of slow, mildly recycled,
accreting pulsars.

}


\maketitle
\titlerunning{Spin and orbit of {\fullsrc}}
\authorrunning{Papitto et al.}

\section{Introduction}

The dense environment of a globular cluster and the 
resulting frequent stellar encounters \citep{Mey97} make the
  production of binary systems hosting a compact object very
  efficient.  Terzan 5 is probably one of the densest and
  metal-richest cluster in our Galaxy \citep{Chn02,Ort07}, as clearly
  indicated by the large number of rotation-powered millisecond
  pulsars discovered there ($\simgt30$, \citealt{Rns05,Hss06}). The
  cluster also contains at least 28 discrete X-ray sources, a
  substantial number of which can be identified as quiescent low-mass
  X-ray binaries \citep[LMXB, ][]{Hnk06}.  According to the recycling
  scenario \citep[see, e.g.][]{BhtvdH91}, the population of
  rotation-powered millisecond pulsars and low-mass X-ray binaries
  (LMXB) share an evolutionary link, because the former are thought to
  be spun up by the accretion of mass and angular momentum in an LMXB.
  Accreting pulsars in LMXB are usually found with periods clustering
  in two distinct groups. So far, 13 sources have been discovered with
  spin periods lower than 10 ms and were therefore termed
  accretion-powered millisecond pulsars \citep[see,
    e.g.,][]{WijvdK98}. However, a smaller number of sources are found
  with longer periods and correspondingly higher estimates of the
  neutron star (NS) magnetic field.

So far, the only bright transient LMXB known in the cluster Terzan 5
was the burster EXO 1745-248 \citep{Mks81}.  The first detection of a
new outburst of a source in this cluster was made on 2010 October
10.365 with {\it INTEGRAL} \citep{Brd10} and was tentatively
attributed to EXO 1745-248.  Follow-up {\it Swift} observations
refined the source position, and a comparison with the position of
sources previously known thanks to {\it Chandra} observations of the
cluster suggested instead a different association \citep{Hnk10, Knn10,
  Poo10}. The X-ray transient is then considered a newly discovered
source and named as {\fullsrc} ({\src} in the following).  A coherent
signal at a period of 90.6 ms was detected thanks to observations
performed with the {\fullrxte} \citep[{\rxte} in the following,
][]{Str10}.  A signal at the same period is present also during the
several bursts that the source shows \citep{Alt10}, while burst
oscillations have never been observed from NS rotating at periods
exceeding a few ms.  A sudden decrease of the flux was tentatively
attributed to an eclipse of the source by the companion
\citep{Str10}. However, eclipses were not observed during subsequent
observations, and the earlier flux decrease was identified with a
lunar occultation \citep[][S10 in the following]{Str10bis}.

Below we present the first analysis of the properties of the coherent
signal emitted by this source, using {\rxte} and {\it Swift}
observations performed between 2010 October 10 and November 6, and
give a refined orbital and timing solution of the pulsar with respect
to those first proposed (\citealt{Ppt10}, P10 in the following, S10).

\section{Observations and data analysis}
\label{sec:obs}

\begin{figure}
\resizebox{\hsize}{!}{\includegraphics{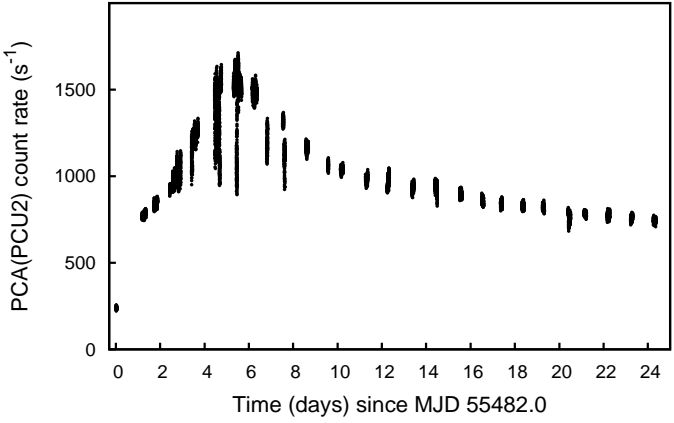}}
\caption{{\src} in the time interval considered here as observed by
  the PCU2 of the PCA aboard {\rxte}.
}
\label{fig1}
\end{figure}

After the source discovery on 2010 October 10.365 \citep{Brd10},
{\rxte} started monitoring the source on MJD 55482.010 (October
13.010; all the epochs reported in this paper are given with regard to
the Barycentric Dynamical Time, TDB, system). We present an analysis
of the observations performed until MJD 55506.359 (ObsId
95437-01-12-01), for a total exposure of 206 ks. In this time
interval, a large number ($>$ 300) of X-ray bursts are observed with a
recurrence time decreasing from $\simgt26$ to $\sim3$ min as the X-ray
flux increases. The analysis of the bursts shown by the source, as
well as of its aperiodic timing properties, will appear in a companion
paper. The 2.5-25 keV light curve recorded by PCU2 of the Proportional
Counter Array (PCA) on-board RXTE, with the burst intervals removed
and background subtracted, is plotted in Fig.\ref{fig1}.  The count
rate increases during the first days of the outburst, reaching a peak
value of $\sim1700$ s$^{-1}$ at MJD 55487.5, and then decreases to a
value of $\sim 800$ s$^{-1}$ with an exponential decay time scale of
$\sim5$ d. Quite complex dipping-like structures also appear
especially between MJD 55485 and 55490 as the flux shows sudden
variations up to 75$\%$ on timescales of $\sim10$ min.

The combined X-ray spectrum observed by the top layer of the PCU2 of
the PCA (2.5--25 keV), and by Cluster A of the High Energy X-ray
Timing Experiment (HEXTE, 22--50 keV), can be well modelled by a sum
of a blackbody and a Comptonized component, which we model with
\texttt{compps} \citep{PouSve96}.  Throughout the observations
considered here, the spectrum softens significantly as the source
evolves towards higher luminosities. The unabsorbed total flux,
extrapolated in the 0.1-100 keV band, rises from $0.47(3)\times
10^{-8}$ to a maximum level of $1.89(4)\times10^{-8}$ erg cm$^{-2}$
s$^{-1}$, observed during the observation of MJD 55487.5. All the
uncertainties on the fluxes given here are quoted at a 90\% confidence
level.

To analyse the spin and orbital properties of the source we consider
data taken by the PCA in Event ($122\mu$s temporal resolution) and
Good Xenon ($1\mu$s temporal resolution) packing modes. We discard 10s
prior, and 100s after the onset of each type-I X-ray burst.  The time
series were also preliminarly barycentred with respect to the solar
system barycentre using the available {\rxte} orbit files and assuming
the best {\it Chandra} estimate of the source position, RA=17$^{h}$
48$^{m}$ 4.831$^{s}$, DEC=-24$^{\circ}$ 46' 48.87'', with an error
circle of 0.06'' (1$\sigma$ confidence level, \citealt{Hnk06,Poo10}).
A coherent signal at a frequency of 11.045(1) Hz [equivalent to a
  period of 90.539(8) ms] is clearly detected in the power spectrum at
a Leahy-normalised power of $\simeq 1.5\times10^4$. A first orbital
solution is obtained modelling the observed modulation caused by the
orbital motion,
\begin{equation}
P(t)= P_0\left\{1+\frac{2\pi x}{P_{orb}} [\cos{m}+e\cos(2m-\omega)]\right\}. 
\end{equation} Here $P_0$ is the
barycentric spin period of the source, $x=a\sin{i}/c$ the semi-major
axis of the NS orbit, $P_{orb}$ the orbital period,
$m=2\pi(t-T*)/P_{orb}$ the mean anomaly, $T^*$ the epoch at which
$m=0$, $e$ the eccentricity and $\omega$ the longitude of the
periastron measured from the ascending node. The periods $P(t)$ are
estimated by performing an epoch-folding search on 1.5 ks long data
segments (for a total of 127) around the periodicity indicated by the
power spectrum. The resulting variance profiles are fitted following
\citet{Leh87} and the uncertainties affecting the period estimates are
evaluated accordingly. The best-fitting orbital solution we obtain
with this technique is shown in the leftmost column of Table
\ref{tab}.

\begin{figure}
\resizebox{\hsize}{!}{\includegraphics{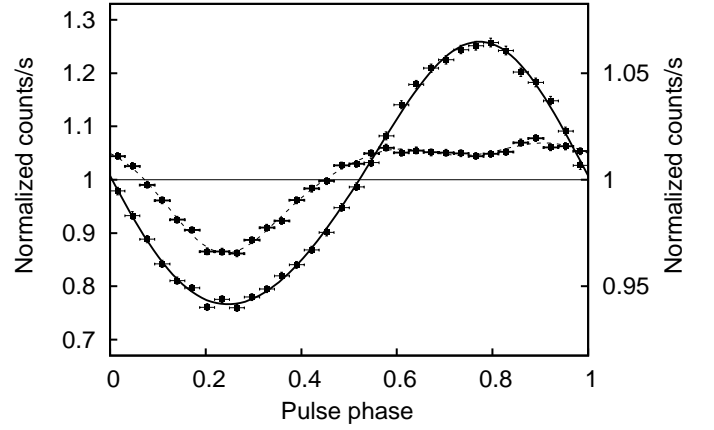}}
\caption{Pulse profiles and best-fitting harmonic decomposition,
  obtained by folding observations performed by {\rxte} on MJD 55482
  (left scale, solid line) and MJD 55483 (right scale, dashed lines)
  around $P_F=90.539645$ ms.  Both profiles are evaluated after
  background subtraction and are normalised to the current average
  flux.}
\label{fig3}
\end{figure}

The time series were then corrected for the orbital motion with these
parameters and folded around the best estimate of the spin period over
300s data segments (for a total of 717). The pulse profiles could
generally be modelled using up to three harmonics. The pulsed fraction
is observed to greatly vary in between the first (MJD 55482.010 to
55482.043) and the other {\rxte} observations. During the former the
pulse fractional amplitude of the first harmonic is very high
[$A_1\simeq0.252(2)$], while the second harmonic is detected at a much
lower amplitude [$A_2=0.016(2)$]. In subsequent observations the
amplitude of the first harmonic drastically decreases to values
between 0.02 and 0.04, whereas the second harmonic amplitude remains
stable and a third harmonic is sometimes requested by the profile
modelling. To show this, we plot in Fig. \ref{fig3} the pulse profiles
calculated over the observations performed during MJD 55482 (solid
line) and MJD 55483 (dashed line), with the latter profile shown at a
magnified scale. The pulsed fraction decrease is evident, as are the
variation of the shape of the peak at rotational phase $\sim0.75$.

To increase the accuracy of our timing solution we model the temporal
evolution of the phases evaluated on the first and second harmonic of
the pulse profiles with the relation:
\begin{equation}
 \label{eq:phases}
\phi(t)=\phi_0+(\nu_0-\nu_{F})\;(t-T_{ref})+\frac{1}{2}{<\dot{\nu}>}(t-T_{ref})^2+R_{orb}(t).
\end{equation}
Here $T_{ref}$ is the reference epoch for the timing solution, $\nu_0$
is the pulsar frequency at the reference epoch, $\nu_F=1/P_F$ is the
folding frequency and $<\dot{\nu}>$ is the average spin frequency
derivative.  The term $R_{orb}(t)$ describes the phase residuals
induced by a difference between the actual orbital parameters, namely
$x$, $P_{orb}$, $T^*$, $e\sin{\omega}$ and $e\cos{\omega}$, and those
used to correct the time series \citep[see e.g.][]{DeeBoyPrv81}. Once
a new set of orbital parameters is found, it is used to correct the
time series, and the resulting phases are modelled again with
Eq.(\ref{eq:phases}). This procedure is iterated until the phase
residuals are normally distributed around zero.

 \begin{table}
\caption{Spin and orbital parameters of {\fullsrc}.\label{tab}} \centering
\begin{tabular}{lrrr}
& Periods  &1$^{st}$ harm. ph. &2$^{nd}$ harm. ph.\\
\hline
$\nu_0\:$-11.044885 ($\mu$Hz) & $<5$ & +0.64(1)& +0.17(1)\\
$<\dot{\nu}>$ ($10^{-12}$ Hz/s) &$<16$ &$1.22(1)$&$1.68(1)$  \\ 
\hline
a$\sin{i}$/c (lt-s)& 2.498(5)&2.4967(3) &2.4973(2)\\ 
P$_{orb}$ (hr) &21.2744(8) &21.2745(1) &21.27454(8)\\ 
T* - 55481.0 (MJD) &0.7805(4)&0.78033(6) &0.78048(4)\\ 
e &$<0.02$ &$<7\times10^{-4}$&$<6\times10^{-4}$\\ 
f($M_1$,$M_2$,i) (M$_{\odot}$) & 0.0213(2)&0.0212587(8)&0.021275(5)\\ 
\hline 
$\chi^2$/dof & 156/121 &7879/662&3287/566\\ 
\hline
\end{tabular}
\tablefoot{ Numbers in parentheses are 1$\sigma$ errors on the last
  significant digit. Upper limits are evaluated at 3$\sigma$
  confidence level. The uncertainties of the timing parameters have
  been scaled by a factor $\sqrt{\chi_r^2}$ to take into account a
  reduced $\chi^2$ of the best-fitting model larger than 1.  The
  orbital solution obtained modelling the Doppler shifts on the pulse
  period is given in the leftmost column, while the best timing
  solution evaluated from the phase evolution of the first and second
  harmonic are given in the central and in the rightmost column,
  respectively. The reference epoch for these timing solutions is MJD
  55483.186. }
\end{table}

\begin{figure}
\resizebox{\hsize}{!}{\includegraphics{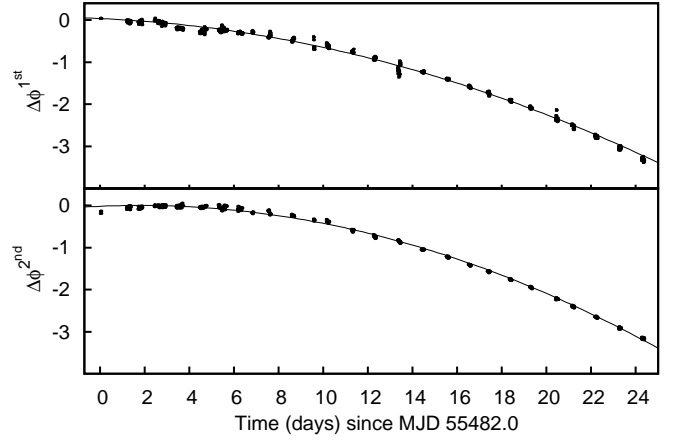}}
\caption{ { Evolution of the first (top) and second (bottom) harmonic
    phases obtained by folding the orbitally corrected time series
    around $P_F=90.539645$ ms (see text). The solid lines show the
    best-fit models from Eq. (\ref{eq:phases}).}}
\label{fig2}
\end{figure}

Although no residual modulation at the orbital period is observed, the
phases of the first harmonic are strongly affected by timing
noise. The reduced $\chi^2$ we obtain modelling their evolution with
Eq.(\ref{eq:phases}) is extremely large ($\simeq 11.9$ over 662
d.o.f.). Such a behaviour is most probably caused by to pulse shape
changes like the one shown in Fig \ref{fig3}.  The second harmonic
phases appear to be less affected by timing noise, resulting in a
reduced $\chi^2_r=5.8$ (566 d.o.f.). We argue that the second harmonic
phases are better fitted with respect to the first harmonic because of
the greater stability of this component \citep[as already observed in
  some accreting millisecond pulsars, see, e.g.,][]{Brd06,Rgg08}.  The
best-fitting parameters calculated over the phase evolution of the
first and second harmonic are quoted in the central and rightmost
column of Table \ref{tab}.  In Fig. \ref{fig2} we show the phases of
both harmonics, when the observations corrected for the orbital motion
of the source are folded around $P_F=90.539645$ ms. The phase
evolution is clearly driven by at least a quadratic component. A
consequence of timing noise is that the spin frequency and its
derivative, estimated over the two harmonic components, are
significantly different. We quote conservatively a value of
$\nu_0=11.0448854(2)$ Hz that overlaps both frequency estimates, and
use a spin frequency derivative between 1.2--1.7$\times10^{-12}$ Hz
s$^{-1}$ in the discussion below.  However, it is worthwhile to note
that the orbital parameters are entirely consistent between the two
harmonic solutions, which supports the reliability of these estimates.
The solution we obtained is entirely compatible with, but more precise
than, those proposed by P10\footnote{There is an offset between the
  values of frequency and epoch of mean longitude quoted by P10 and
  those presented here, as theirs were not referred to the TDB
  reference system.}  and S10. Given the accuracy of the source
position considered here (0.06''), the systematic uncertainties
introduced by the position indetermination on the measured values of
spin frequency and of its derivative \citep[e.g.][]{Brd07} are
$\sigma_{pos,\nu}\simlt3\times10^{-10}$ Hz and
$\sigma_{pos,\dot{\nu}}\simlt6\times10^{-17}$ Hz s$^{-1}$,
respectively.  Finally, as the cluster moves towards the solar system
at a velocity of $85\pm10$ km s$^{-1}$ \citep{Fer09}, the measured
value of the spin frequency is affected by a systematic offset of
$\sim +3\times10^{-3}$ Hz, though this is unimportant when making
  conclusions about the source properties.

  In order to extend the range of fluxes at which the source was
  observed and pulsations were detected, we also analysed three {\it
    Swift} observations (Obs. 00031841002, 00031841003 and
  00031841004) in which the XRT observed in Windowed Timing (WT) mode,
  with a temporal resolution of 1.7ms. The {\it Swift} XRT started
  monitoring the source on MJD 55479.737, more than two days before
  {\rxte}, with the first observation in WT mode starting at MJD
  55479.802 for 2 ks. After applying barycentric corrections for the
  satellite orbit, then correcting for the source orbital motion and
  selecting photons from a 50 pixel wide box around the source
  position, a pulsation is clearly detected at a period of
  $P_S=90.5395(2)$ ms by means of an epoch-folding search. The XRT
  signal is particularly strong and consistent with that seen by
  {\rxte} during its first observation, with a pulse profile modelled
  by a sinusoid of amplitude 0.23(2).  Pulsations were also searched
  for in the subsequent $\sim1$ ks long XRT WT observations starting
  on MJD 55484.767 and 55485.357. Only a weak signal was detected in
  the latter at a fractional amplitude of 0.018(4), still compatible
  with that seen by {\rxte} at those later times.

\section{Discussion and conclusions}

We reported on the spin and orbital properties of the newly discovered
accreting pulsar, {\fullsrc}. Its 90.6 ms period makes it the first
confirmed accreting pulsar in the range 10--100 ms.  Pulsations were
detected in all observations performed by {\rxte}, as well as in two
out of the three {\it Swift} observations performed in WT mode
presented here.The pulsed fraction is observed to drastically change
on a timescale of $\simlt 1$ d, after the {\rxte} observation
performed on MJD 55482. While previously both {\it Swift} and {\rxte}
observations revealed a strong signal dominated by a first harmonic
component of fractional amplitude as large as $0.25$, later
observations at higher fluxes performed by both satellites never
detected an amplitude $\simgt 0.03$. Simultaneously the pulse shape
changes and becomes more complex.  This behaviour is suggestive of a
change of the geometrical properties of the flow in the accretion
columns above the NS hot spots.

Because pulsations are detected throughout the observations shown
here, an estimate of the NS magnetic field strength can be made.  For
accretion to proceed and pulsations to appear in the X-ray light
curve, the inner disc radius, $R_{in}$, has to lie in between the NS
radius, R$_{NS}$, and the corotation radius, $R_{C}$, defined as the
distance from the NS at which the velocity of the magnetosphere equals
the Keplerian velocity of the matter in the disc. For a larger
accretion radius, accretion would be inhibited or severely reduced by
the onset of a centrifugal barrier. For a pulsar spinning at 90.6 ms,
$R_C=(GMP^2/4\pi^2)^{1/3}=338 m_{1.4}^{1/3}$ km, where $m_{1.4}$ is
the NS mass in units of 1.4 M$_{\odot}$. Defining the inner disc
radius in terms of the pressure equilibrium between the disc and the
magnetosphere, one obtains $R_{in}\simeq 160 \;
m_{1.4}^{1/7}\;R_6^{-2/7}\;L_{37}^{-2/7}\mu_{28}^{4/7}$ km
\citep{Brd01}, where $R_6$ is the NS radius in units of 10 km,
$L_{37}$ the accretion luminosity in units of $10^{37}$ erg s$^{-1}$,
and $\mu_{28}$ the magnetic dipole moment of the NS in units of
$10^{28}$ G cm$^{3}$.  Extrapolating the fluxes observed by {\rxte} to
the 0.1--100 keV band, and assuming a distance of $d=5.5\pm0.9$ kpc to
Terzan 5 \citep{Ort07}, we estimate a maximum and minimum bolometric
luminosities of $1.7(1)\times10^{37}$ d$_{5.5}^2$ erg s$^{-1}$ and
$6.8(1)\times10^{37}$ d$_{5.5}^2$ erg s$^{-1}$, during the time
covered by the observations considered here. Assuming that the X-ray
luminosity is a good tracer of the accretion power and imposing
$R_{NS}<R_{in}\simlt R_{C}$, we obtain
\begin{equation}
0.02\: \: m_{1.4}^{-1/4}\: R_{6}^{9/4}\: d_{5.5} \simlt \mu_{28} \simlt 4.8 \:\: m_{1.4}^{1/3}\: R_{6}^{1/2}\: d_{5.5}.
\end{equation}
The upper limit on the magnetic dipole can be reduced considering that
pulsations are detected also in a Swift observation taking place
$\sim2$ d earlier than the first {\rxte} observation. \citet{Bzz10}
estimated the source flux in that observation as
$4.5(2)\times10^{-10}$ erg cm$^{-2}$ s$^{-1}$ (1--10 keV). This value
is a factor $\sim4$ lower than the value obtained extrapolating the
spectrum of the first {\rxte} observation in the same energy
band. Assuming that this ratio holds also for the bolometric
luminosity of the source, we get to an upper limit on the magnetic
dipole moment of $\simeq 2.4\times10^{28}$ G cm$^{-3}$.  The limits
thus derived translate to a magnetic surface flux density between
$\sim 2\times10^8$ and $\sim 2.4\times10^{10}$ G. The upper bound of
this interval can be overestimated because the exact flux at which the
pulsations appeared is unknown at present. Monitoring the presence of
coherent pulsations as a function of the flux when the source fades
will probably allow us to  derive a tighter constraint.
\citet{Alt10bis} have also reported the presence of a kHz QPO at
$\sim815$ Hz (10--50 keV) during the {\rxte} observations performed on
MJD 55487.  Under the hypothesis that this feature originates in the
innermost part of the accretion disc, it indicates an inner disc
radius $R_{in}\simlt \:20 \:m_{1.4}^{1/3}$ km. As the luminosity we
estimated during that day is $6.8(1)\times10^{37}$ d$_{5.5}^2$ erg
s$^{-1}$, this would imply a magnetic field $\simlt7\times 10^8$
d$_{5.5}$ G if the disc is truncated at the magnetospheric radius.

Despite the presence of timing noise, the analysis of the phase
evolution over the $\sim24$ d time interval presented here clearly
indicates the need for a quadratic component to model these phases.
Interpreting this component as a tracer of the NS spin evolution, we
thus conclude that the source spins up while accreting.  Values of the
spin-up rate between 1.2 and 1.7$\times10^{-12}$ Hz s$^{-1}$ are
found, depending on the harmonic considered. This discrepancy is
probably due to the effect of timing noise. These values are
compatible with those expected for a NS accreting the Keplerian disc
matter angular momentum given the observed luminosity,
$\dot{\nu}\simeq1.5\times10^{-12}$ $(L_{37}/5)$ $(R_{in}/70km)^{1/2}$
$I_{45}^{-1}$ $R_6$ $m_{1.4}$ Hz s$^{-1}$. Here $I_{45}$ is the NS
moment of inertia in units of $10^{45}$ g cm$^2$.  The observed spin
period and the magnetic field we estimated place this source between
the population of ``classical'' ($B\simgt 10^{11}$G, $P\simgt 0.1$ s)
and millisecond ($B\simeq 10^8$--$10^9$ G, $P\simeq 1.5$--$10$ms)
rotation-powered pulsars. The observation of a significant spin-up at
rates compatible with those predicted by the recycling scenario
further supports the identification of this source as a slow, mildly
recycled pulsar. We note that the only other two accreting pulsars
with similar, though significantly different parameters, are GRO
J1744-28 ($P_S=467ms$, $B\simeq2.4\times10^{11}$ G, \citealt{Cui97}),
and 2A 1822-371 ($P_S=590$ ms, $B\simlt 10^{11}$ G, \citealt{Jnk01}).

The orbital parameters we measured for the NS in {\src} allow us to
derive constraints on the nature of its companion star. With a mass
function of $f(M_2;M_1,i)\simeq 0.02$ M$_{\odot}$, a minimum mass for
the companion can be estimated to be as low as 0.41 M$_{\odot}$ for an
inclination of 90$^{\circ}$, and an NS mass of 1.4
M$_{\odot}$. Since the source shows no eclipses, the inclination is
most probably $\simlt80^{\circ}$, { and} the lower limit increases to
$m_2=0.16+0.26 m_{1.4}$, where $m_2$ is the mass of the companion star
in solar units.  An upper limit can be obtained if the companion star
{ is assumed not to} overfill its Roche lobe.  Using the relation
given by \citet{Egg83} and the third Kepler law to relate the Roche
Lobe radius to the orbital period and to the companion mass,
$R_{L2}\simeq
0.55(GM_{\odot})^{1/3}(P_{orb}/2\pi)^{2/3}m_{1.4}\:q^{2/3}(1+q)^{1/3}/[0.6q^{2/3}+\log(1+q^{1/3})]$,
where $q=M_2/M_1$, and assuming the companion star follows a main
sequence mass-radius relation, $R_2/R_{\odot}\approx(M_2/M_{\odot})$,
yields a maximum mass of 2.75 M$_{\odot}$ for the companion when
$m_{1.4}=1$. This upper limit is indeed higher than the maximum mass
expected for a main sequence star belonging to one of the two stellar
populations found by \citet{Fer09} in Terzan 5. One has in fact
$m_2\simlt0.95$ if the companion of {\src} belongs to the older
population \citep[t=10 Gyr,][]{Dan10}, while $m_2\simlt 1.2$ and
$m_2\simlt 1.5$ if it belongs to a younger population of 6 and 4 Gyr,
respectively (D'Antona, priv. comm.). We conclude that a
reasonable upper limit for the companion-star mass is 1.5 M$_{\odot}$,
possibly a main sequence or a slightly evolved star.

\begin{acknowledgements}
 This work is supported by the Italian Space Agency, ASI-INAF
 I/088/06/0 contract for High Energy Astrophysics, as well as by the
 operating program of Regione Sardegna (European Social Fund
 2007-2013), L.R.7/2007, ``Promotion of scientific research and
 technological innovation in Sardinia''. We thank F. D'Antona for
 providing the mass estimates of main sequence stars in Terzan 5, and
 the referee for the prompt reply and useful comments.
\end{acknowledgements}

\bibliographystyle{aa}
\bibliography{biblio}

\end{document}